%% file: main.tex
\documentclass[lettersize,journal]{IEEEtran}
\usepackage{amsmath,amsfonts}
\usepackage{array}
\usepackage[caption=false,font=normalsize,labelfont=sf,textfont=sf]{subfig}
\usepackage{textcomp}
\usepackage{stfloats}
\usepackage{setspace}
\usepackage{url}
\usepackage{verbatim}
\usepackage{graphicx}
\def\BibTeX{{\rm B\kern-.05em{\sc i\kern-.025em b}\kern-.08em
    T\kern-.1667em\lower.7ex\hbox{E}\kern-.125emX}}
\usepackage{balance}

\usepackage{comment}
\usepackage{paralist}
\usepackage{bm}
\usepackage{pgfplots}
\usetikzlibrary{pgfplots.dateplot}

\usepackage{algpseudocode}
\usepackage[ruled,linesnumbered]{algorithm2e}
\usepackage{booktabs}

\usepackage{flushend}
\usepackage[english]{babel}
\usepackage[latin1]{inputenc}
\usepackage{multirow} 
\usepackage{mathrsfs}
\usepackage{graphicx}

\usepackage{amssymb}
\usepackage{amsfonts}
\usepackage{url}
\usepackage{longtable}
\usepackage{rotating}
\usepackage{mathrsfs}
\usepackage{hyperref}
\usepackage{pgfplots}
\usetikzlibrary{pgfplots.dateplot}
\usepackage{filecontents}

\usepackage{colortbl}
\usepackage{threeparttable}
\usepackage{makecell}

\definecolor{tblue}{RGB}{51,139,180}
\definecolor{torange}{RGB}{235,157,14}
\definecolor{tgreen}{RGB}{44,170,44}
\definecolor{tred}{RGB}{214,99,40}
\definecolor{tpurple}{RGB}{148,133,189}
\definecolor{LightCyan}{rgb}{0.88,1,1}
\definecolor{hidden-draw}{RGB}{40,158,106}
\definecolor{hidden-pink}{RGB}{255,245,247}
\definecolor{lightred}{RGB}{255, 204, 204}
\definecolor{lightgreen}{RGB}{224, 255, 225}
\definecolor{lightyellow}{RGB}{255, 241, 224}
\definecolor{lightblue}{rgb}{0.901, 1, 1}
\definecolor{lightpurple}{RGB}{225, 225, 255}
\definecolor{purple}{RGB}{202, 206, 218}
\definecolor{lightgray}{gray}{0.9}

\newcommand{\hide}[1]{} 

\def\model{\textbf{M$^2$Rec}}
\def\fmodel{AFFM}

\begin{document}

\title{M$^2$Rec: \underline{M}ulti-scale \underline{M}amba for Efficient Sequential \underline{Rec}ommendation}

\author{Qianru Zhang, Liang Qu, Honggang Wen, Dong Huang, Siu-Ming Yiu\IEEEauthorrefmark{5}, \\Nguyen Quoc Viet Hung, Hongzhi Yin\IEEEauthorrefmark{5}
\IEEEcompsocitemizethanks{
\IEEEcompsocthanksitem
\IEEEauthorrefmark{5}Corresponding author.
\IEEEcompsocthanksitem
Q. Zhang, H. Wen, D. Huang and S.M. Yiu are from the school of computing and data science, the University of Hong Kong. E-mail: \{qrzhang,smyiu\}@cs.hku.hk, 
whgtytyg@gmail.com, u3008427@connect.hku.hk.
\IEEEcompsocthanksitem 
L. Qu works at the School of Business and Law, Edith Cowan University. E-mail: l.qu@ecu.edu.au.
\IEEEcompsocthanksitem 
Q. Hung works at School of Information and Communication Technology, Griffith University. E-mail: henry.nguyen@griffith.edu.au.
\IEEEcompsocthanksitem 
H. Yin works at School of Electrical Engineering and computer science, the University of Queensland. E-mail: db.hongzhi@gmail.com.
}
}

\maketitle

\input{abstract}

\pagenumbering{arabic}
\setcounter{page}{1}


\IEEEdisplaynontitleabstractindextext

\IEEEpeerreviewmaketitle

\input{intro}

\input{relate}

\input{solution}

\input{eval}

\input{model}

\input{conclusion}


\bibliographystyle{IEEEtran}
\bibliography{main}

\end{document}

%% file: abstract.tex
\begin{abstract}

Sequential recommendation systems aim to predict users' next preferences based on their interaction histories, but existing approaches face critical limitations in efficiency and multi-scale pattern recognition. While Transformer-based methods struggle with quadratic computational complexity, recent Mamba-based models improve efficiency but fail to capture periodic user behaviors, leverage rich semantic information, or effectively fuse multimodal features. To address these challenges, we propose \model, a novel sequential recommendation framework that integrates multi-scale Mamba with Fourier analysis, Large Language Models (LLMs), and adaptive gating. First, we enhance Mamba with Fast Fourier Transform (FFT) to explicitly model periodic patterns in the frequency domain, separating meaningful trends from noise. Second, we incorporate LLM-based text embeddings to enrich sparse interaction data with semantic context from item descriptions. Finally, we introduce a learnable gate mechanism to dynamically balance temporal (Mamba), frequency (FFT), and semantic (LLM) features, ensuring harmonious multimodal fusion. Extensive experiments demonstrate that \model\ achieves state-of-the-art performance, improving Hit Rate@10 by 3.2\% over existing Mamba-based models while maintaining 20\% faster inference than Transformer baselines. Our results highlight the effectiveness of combining frequency analysis, semantic understanding, and adaptive fusion for sequential recommendation. Code and datasets are available at: https://anonymous.4open.science/r/M2Rec.
\end{abstract}

%% file: intro.tex
\section{Introduction}
\label{sec:intro}

\begin{figure*}
\centering
\includegraphics[width=0.75\linewidth, height=0.35\linewidth]{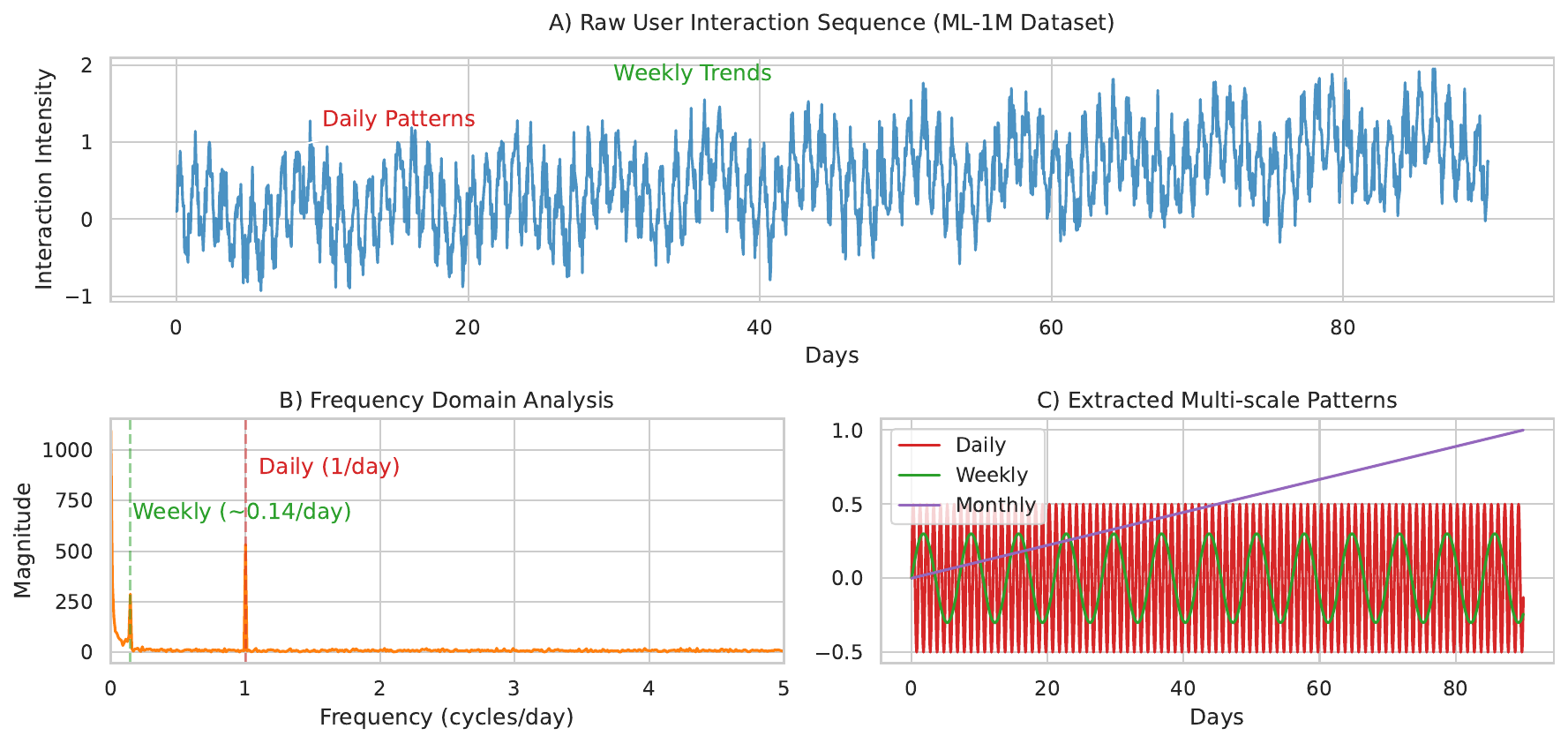}
\vspace{-0.15in}
\caption{The diagram illustrates the multi-scale temporal patterns in user interactions from the MovieLens 1M (ML-1M) dataset.}
\vspace{-0.1in}
\label{fig:intro_scale}
\end{figure*}

\IEEEPARstart{S}{equential} recommendation systems aim to predict users' next preferred items based on their interaction histories~\cite{wang2019sequential,boka2024survey,quadrana2018sequence}, playing crucial roles in e-commerce, streaming platforms~\cite{chang2021sequential,li2021lightweight,huang2018improving}, and other personalized services~\cite{tang2018personalized,wu2024personalized,zhang2024recgpt}. While Transformer-based models~\cite{vaswani2017attention,de2021transformers4rec,fan2021continuous,yang2024sequential} like BERT4Rec~\cite{sun2019bert4rec}, SSE-PT~\cite{wu2020sse} have demonstrated strong performance in capturing long-range dependencies, their quadratic computational complexity $\mathcal{O}(L^2)$ severely limits scalability for long sequences. The recent emergence of Mamba, with its efficient $\mathcal{O}(Lw)$ complexity, presents a promising alternative, but three critical challenges remain unaddressed in applying Mamba~\cite{gu2023mamba} to sequential recommendation tasks.

First, user interactions~\cite{liu2024mamba4rec,yang2024uncovering,zhang2024matrrec,qu2024ssd4rec,su2024mlsa4rec} often exhibit multi-scale periodic patterns - daily browsing habits, weekly purchase cycles, or seasonal preferences - that are typically obscured by noise in raw sequence data. While Mamba effectively captures local temporal dynamics, it lacks explicit mechanisms to identify and amplify these meaningful periodic patterns while filtering out noise. Second, existing Mamba-based recommendation approaches~\cite{liu2024mamba4rec,yang2024uncovering} rely solely on sparse ID-based features, ignoring the rich semantic information contained in item descriptions and user reviews that could provide valuable insights into user preferences. Third, simply concatenating features from different modalities - temporal patterns from Mamba, frequency components from Fourier analysis, and semantic embeddings from language models - often leads to conflicting signals and suboptimal performance due to poor feature alignment.

To address these challenges, we propose \model, an innovative framework that strategically combines three key technical components. First, we enhance Mamba with Fourier Transform capabilities, enabling explicit modeling of periodic patterns in the frequency domain. This FFT-enhanced module~\cite{zhang2025frequency,sneddon1995fourier} decomposes interaction sequences into their frequency components, allowing the model to separately analyze and weight different temporal scales - rapidly identifying daily patterns while also capturing longer weekly or monthly cycles. Crucially, this frequency-domain analysis provides a natural mechanism for noise reduction by attenuating high-frequency components that typically correspond to random variations rather than genuine user preferences. For example, Figure ~\ref{fig:intro_scale} illustrates the multi-scale temporal patterns in user interactions from the ML-1M dataset. Panel A shows the raw interaction sequence where three characteristic patterns coexist: (1) high-frequency daily cycles (red arrows) reflecting circadian rhythms in user activity, (2) medium-frequency weekly variations (green arrows) indicating weekend vs. weekday behavioral differences, and (3) low-frequency monthly drift (purple arrows) representing evolving user preferences. Panel B demonstrates the corresponding frequency spectrum analysis via Fast Fourier Transform (FFT), where dominant peaks at 1 cycle/day (red dashed line) and $\sim$0.14 cycles/day (green dashed line) quantitatively confirm the presence of daily and weekly patterns. Panel C decomposes the original signal into these interpretable components. This multi-scale analysis motivates our approach to sequential recommendation, where explicitly modeling such temporal structures can improve prediction accuracy while maintaining computational efficiency.

Second, we integrate Large Language Models (LLMs)~\cite{harte2023leveraging,boz2024improving} to extract rich semantic embeddings from item descriptions and user reviews. While traditional recommendation systems rely on sparse item IDs, these text embeddings capture nuanced product characteristics and user sentiments that are essential for understanding preferences. For instance, LLM embeddings can distinguish between different shades of meaning in product descriptions or identify subtle sentiment cues in reviews - information that would be lost in a pure ID-based approach.

Third, we introduce a novel gated fusion mechanism~\cite{li2018self} to intelligently combine these diverse information sources. Rather than simply concatenating features, the gate learns to dynamically adjust the contribution of each modality (temporal patterns from Mamba, periodic signals from FFT, and semantic information from LLMs) based on context. This allows the model to, for example, emphasize semantic signals when recommending books while relying more on temporal patterns for frequently purchased grocery items.

The combination of these components in \model\ yields several important advantages. The FFT enhancement allows explicit modeling of multi-scale patterns that are difficult to capture through temporal analysis alone. LLM integration addresses the sparsity problem inherent in ID-based approaches while providing rich semantic understanding. The gated fusion mechanism ensures these different information streams complement rather than conflict with each other. Importantly, all these enhancements maintain Mamba's core efficiency advantages, making \model\ scalable to long sequences and large catalogs.

We summarize our contributions as follows:

\begin{compactitem}
\item \textbf{An Efficient Unified Approach for Sequential Recommendation}. In this paper, we introduce a cohesive and efficient framework tailored for multi-scale Sequential Recommendation tasks.
\item \textbf{Utilizing the Adaptive Fourier Transform, LLMs, and Gate Mechanism, our enhanced Mamba-based model excels in capturing periodic patterns and reducing noise, leveraging rich semantic information, and enhancing fusion capabilities across diverse modal features.} By integrating the Fourier Transform and incorporating LLMs via the Gate Mechanism, our novel \model\ effectively captures intricate multi-scale periodic patterns of user patterns unique to sequential recommendation tasks and extracts valuable semantics from LLMs. Additionally, our approach facilitates the fusion of diverse resources through the Gate mechanism, enhancing the model's ability to fuse multimodal information, ultimately leading to performance enhancement.
\item \textbf{Comprehensive Experiments.} Through rigorous evaluations on sequential recommendation datasets and comparisons with robust benchmarks such as transformer-based models and other Mamba-inspired architectures, our \model\ attains cutting-edge performance levels, establishing itself as a leader in the field. Our code and data can be accessed at \href{https://anonymous.4open.science/r/M2ReC-31DF/README.md}{https://anonymous.4open.science/r/M2Rec}.
\end{compactitem}

%% file: relate.tex
\section{Preliminaries}

\begin{figure*}[htb!]
\centering
\includegraphics[width=0.92\linewidth, height=0.25\linewidth]{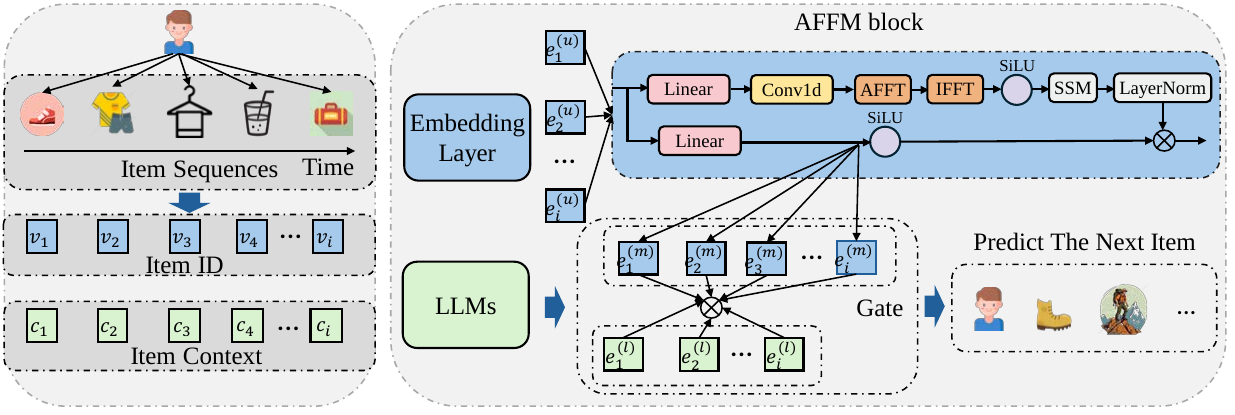}
\vspace{-0.15in}
\caption{The diagram illustrates the configuration of the \model. The left part represents the input data, comprising item sequences with item IDs and item contexts. On the right side, the diagram features the AFFM block and the LLMs responsible for generating embeddings that represent the items.}
\vspace{-0.1in}
\label{fig:fram}
\end{figure*}

\subsection{Fourier Transform}
The Fourier Transform is a powerful mathematical tool used to analyze the frequency components of a signal. It decomposes a time-domain signal into its constituent sinusoidal components, revealing the frequencies and amplitudes present in the signal. The Discrete Fourier Transform (DFT) is the discrete version of the Fourier transform. The DFT of a discrete-time signal $x[n]$ of length $N$ is given by:
\begin{equation}
\label{eq:four1}
X[k] = \sum_{n=0}^{N-1} x[n] e^{-j\frac{2\pi}{N}kn}; ~~~j = \sqrt{-1}
\end{equation}
where $X[k]$ represents the discrete spectrum of the signal and $k$ is the discrete frequency index. And $j$ denotes the imaginary part.

For discrete-time signals, the inverse DFT is shown as follows: 
\begin{equation}
\label{eq:inverse_four1}
x[n] = \frac{1}{N} \sum_{k=0}^{N-1} X[k] e^{j\frac{2\pi}{N}kn};~~~j=\sqrt{-1}
\end{equation}

The Fast Fourier Transform (FFT) is an efficient algorithm used to compute the Discrete Fourier Transform (DFT) of a sequence, which is a fundamental operation in signal processing and many other fields. The FFT greatly accelerates the computation of the DFT compared to traditional methods, making it essential in various applications where the transformation of signals from the time domain to the frequency domain is required.

The FFT reduces the computational complexity of the DFT from $\mathcal{O}(N^2)$ to $\mathcal{O}(N \log N)$, where $N$ is the number of samples in the input sequence. This drastic improvement in efficiency has made the FFT a cornerstone in digital signal processing, telecommunications, audio processing, image processing, and many other scientific and engineering disciplines.

\subsection{State Space Models.} State-space models (SSMs) constitute an elegant framework for representing linear time-invariant (LTI) systems. These models intricately map continuous input signals, denoted as $\mathbf{x}(t)$, to corresponding output signals, $\mathbf{y}(t)$, through the mediation of a latent, hidden state, $\mathbf{h}(t)$. The temporal evolution of this hidden state is elegantly captured by a system of ordinary differential equations (ODEs):
\begin{equation}
\begin{aligned}
\label{eq:ssm1}
\mathbf{h}'(t) = \mathbf{A}\mathbf{h}(t) + \mathbf{B}\mathbf{x}(t)
\end{aligned}
\end{equation}
Here, $\mathbf{A}$ embodies the state transition matrix, while $\mathbf{B}$ and $\mathbf{C}$ function as projection matrices. The S4 and Mamba models exemplify discrete-time state-space models (SSMs). These models leverage a timestep parameter, $\Delta$, and a sophisticated discretization technique-such as the venerable Euler method or the robust Zero-Order Hold (ZOH)-to approximate the continuous-time state transition matrix, $\mathbf{A}$, and input matrix, $\mathbf{B}$, with their discrete-time counterparts, $\bar{\mathbf{A}}$ and $\bar{\mathbf{B}}$ respectively. This discretization process ingeniously adapts the continuous-time SSM formulation, facilitating efficient numerical computation and streamlined implementation. The transformation is achieved via:
\begin{equation}
\begin{aligned}
\bar{\mathbf{A}} &= \exp(\Delta \mathbf{A}) \
\bar{\mathbf{B}} &= \Delta \mathbf{A}^{-1}\exp(\Delta \mathbf{A}) \cdot \Delta \mathbf{B}
\end{aligned}
\end{equation}
The SSM framework discretizes continuous signals into temporally sequenced data points, employing a uniform time interval of $\Delta$. The discrete-time model is expressed as:
\begin{equation}
\begin{aligned}
\label{eq:ssm2}
\mathbf{y}(t) = \mathbf{C}\mathbf{h}(t); \quad \mathbf{h}_t = \mathbf{A}\mathbf{h}_{t-1} + \mathbf{B}\mathbf{x}_t
\end{aligned}
\end{equation}
where $\mathbf{y}(t)$ represents the measurement vector at time $t$, and $\mathbf{C}$ serves as the observation matrix. $\mathbf{h}_t$ and $\mathbf{x}_t$ denote the state vector and input vector, respectively, at time $t$.

During the training phase, the model computes the output using a global convolution:
\begin{equation}
\label{eq:ssm_conv}
\bar{\mathbf{K}} = (\mathbf{C}\bar{\mathbf{B}},\mathbf{C}\bar{\mathbf{A}}\bar{\mathbf{B}},..., \mathbf{C}\bar{\mathbf{A}}^{T-1}\bar{\mathbf{B}}); \quad \mathbf{y} = \mathbf{x} \ast \bar{\mathbf{K}}
\end{equation}
The SSM framework's completeness hinges on specifying initial conditions: $\mathbf{x}(0) = \mathbf{x}_0$, where $\mathbf{x}_0$ represents the initial state vector.

\subsection{Mamba}
\label{sec:ini_mamba}
Mamba~\cite{gu2023mamba} offers a distinct approach compared to the multi-head attention mechanism employed in Transformers, utilizing a selective SSM for capturing feature interactions. The fundamental idea behind Mamba is to map the input sequence $\mathbf{X} = (x_1, x_2, \ldots, x_L)$ to the output $\mathbf{Y}$ through a hidden state $h(i)$, which acts as a linear time-invariant system. In our methodology, we leverage Mamba to model the input sequence $x_i \in \mathbb{R}^{|\mathcal{I}|}$, where $|\mathcal{I}|$ represents the number of correlated items in the sequential recommendation datasets. The process of vanilla Mamba is shown:
\begin{equation}
\begin{aligned}
\bar{\mathbf{A}} &= \exp(\Delta \mathbf{A}) ;~~~\bar{\mathbf{B}} = \Delta \mathbf{A}^{-1}\exp(\Delta \mathbf{A}) \cdot \Delta \mathbf{B}\\
y_i&=\text{SSM}(\bar{\mathbf{A}}, \bar{\mathbf{B}}, \mathbf{C})(x_i)
\end{aligned}
\end{equation}
Here, $x_i \in \mathbb{R}^V$, while $\mathbf{B}$ and $\mathbf{C}$ are learnable matrices of dimensions $(N \times V)$ and $(N \times V)$, respectively. The matrix $\mathbf{A}$ has dimensions $(N \times N)$. The discretization process involves utilizing $\Delta$ to facilitate more efficient calculations and enable the handling of sequential data. Specifically, we obtain $\bar{\mathbf{A}}$ by exponentiating $\Delta \mathbf{A}$, and $\bar{\mathbf{B}}$ is determined using the inverse of $\Delta \mathbf{A}$ and the matrix exponential of $\Delta \mathbf{A}$ multiplied by $\Delta \mathbf{B}$.

\subsection{Problem Definition}
We provide the sequential recommendation problem here. We assume the set of items, denoted as $\mathcal{V} = \left\{v_1, ...,v_{|\mathcal{V}|}\right\}$ and the set of users, denoted as $\mathcal{U} = \left\{u_1, ...,u_{|\mathcal{U}|}\right\}$, here $|\mathcal{V}|$ and $|\mathcal{U}|$ denote the number of items and users. The interaction sequence ranked by time order of user $u \in \mathcal{U}$ is denoted as $\mathcal{X}_{u} = \left\{v^{(u)}_1, ..., v^{(u)}_T\right\}$. Here $v^{(u)}_i \in \mathcal{V}$. Let $\mathcal{C}$ denote the context information set like item comments, etc. Then $c_v$ denotes the comments or reviews information of item $v$. 
\textbf{The task:} we aim to forecast the next item $v_{T+1}$ user $u$ might engage with, utilizing the sequence $\mathcal{X}_{u}$ of interacted items and the content information associated with each.

%% file: solution.tex
\section{Methodology}
\subsection{Overview}

This section explores the technical architecture of our model, \model, which enhances Mamba's capabilities in managing multi-scale periodicities and reducing noise. We start by integrating an embedding layer for nuanced item representations, as detailed in \textbf{Section~\ref{sec:emb}}. Utilizing Large Language Models (LLMs), we develop item embedding vectors that capture item descriptions (\textbf{Section~\ref{sec:llm}}). We then examine the adaptive Fast Fourier Transform (AFFT)-enhanced Mamba encoder, which captures user behavior patterns (\textbf{Section~\ref{sec:adfft}}). The fusion of LLM-derived embeddings and the AFFT encoder is achieved through a sophisticated gate mechanism and optimization strategy (\textbf{Section~\ref{sec:fusion}}). Lastly, we discuss the computational intricacies of the model (\textbf{Section~\ref{sec:complexity}}), shown in Figure~\ref{fig:fram}.

\subsection{Embedding Layer}
\label{sec:emb}
In alignment with prior research~\cite{liu2024mamba4rec}, an embedding layer is incorporated to project item identifiers into a high-dimensional vector space. This layer employs a trainable embedding matrix, denoted as $\mathbf{E}^{(1)} \in \mathbb{R}^{|\mathcal{V}|\times d}$, where $|\mathcal{V}|$ represents the total number of items and $d$ signifies the dimensionality of the embedding vectors.  Applying this embedding layer to the input sequence $\mathcal{X}_u$ yields a set of item embeddings, $\mathbf{E}^{(u)}$. This transformation is formally expressed as:
\begin{equation}
    \begin{aligned}
    \label{eq:item_emb}
        \mathbf{E}^{(u)} = \text{Embedding Layer}(\mathcal{X}_u)
    \end{aligned}
\end{equation}

This equation concisely summarizes the process of mapping the input sequence of item IDs into their corresponding dense vector representations within the embedding space.

\begin{table}[]
\setlength{\tabcolsep}{3.5pt}
\caption{Frequently used notations}
\label{tab:symbols}
\vspace{-0.05in}
\small
\resizebox{0.95\linewidth}{!}{
\begin{tabular}{c|c}
\toprule
\textbf{Notations} & \multicolumn{1}{l}{\textbf{Description}} \\ \midrule
 $\mathcal{V}$         & \multicolumn{1}{l}{The set of items}            \\ 
 $\mathcal{U}$       & \multicolumn{1}{l}{The set of users}            \\ 
 $\mathcal{X}$   & \multicolumn{1}{l}{The user-item interaction sequence}\\ 
  $\mathcal{X}_u$   & \multicolumn{1}{l}{The sequence of the user $u$}\\ 
 $\mathcal{C}$ &\multicolumn{1}{l}{The context information set}\\
 $\mathbf{T}$ &\multicolumn{1}{l}{The item sequence}\\
 $\mathbf{E}^{(1)}$ &\multicolumn{1}{l}{Trainable Embedding matrix of the embedding layer}\\
 $\mathbf{E}^{(u)}$ &\multicolumn{1}{l}{Item embeddings of the embedding layer}\\
 $\mathbf{E}^{(l)}$ &\multicolumn{1}{l}{Item embeddings of LLMs}\\
 $\mathbf{X}_u[k]$ &\multicolumn{1}{l}{The transformed sequence}\\
 $d_f$ &\multicolumn{1}{l}{The sequence length in the frequency domain}\\
 $k$ &\multicolumn{1}{l}{The frequency index}\\
 $j$ &\multicolumn{1}{l}{The imaginary unit}\\
 $\mathbf{X}^{m}_u[k]$ &\multicolumn{1}{l}{The filtered sequence in frequency domain}\\
 $M$ &\multicolumn{1}{l}{The binary mask}\\
 $\theta$ &\multicolumn{1}{l}{The threshold of high frequency}\\
 $\tilde{\mathbf{X}}_u[t]$ &\multicolumn{1}{l}{The time-domain representation}\\
 $\mathcal{K}$ &\multicolumn{1}{l}{The Kernel}\\
 $\mathcal{W}$ &\multicolumn{1}{l}{The Fourier transform of $\mathcal{K}$}\\
 $\mathcal{F}$ &\multicolumn{1}{l}{FFT}\\
 $\mathcal{F}^{-1}$ &\multicolumn{1}{l}{Inverse FFT}\\
 $h_t$ &\multicolumn{1}{l}{The hidden state at $t$ time step}\\
 $\mathcal{H}$ &\multicolumn{1}{l}{The hidden state of a sequence}\\
 $\mathcal{E}$ &\multicolumn{1}{l}{The embedding matrix of the \fmodel module}\\
 $\hat{\mathcal{Y}}$ &\multicolumn{1}{l}{The predicted set of items for users}\\
 $\mathcal{L}$ &\multicolumn{1}{l}{The loss function}\\\bottomrule
\end{tabular}}
\vspace{-0.15in}
\end{table}

\subsection{LLM Augmented Embedding}
\label{sec:llm}
In this section, our objective is to generate embeddings using Large Language Models. After evaluating various models, we found that bge-large-en-v1.5~\cite{chen2024bge} exhibited the best performance in terms of representation ability. Therefore, we decided to utilize this model to optimize the embedding of item sequences. To begin, we describe each item by its name, category, and description, represented as $c_{v} = \{name, category, description\}$. Subsequently, we input this item representation into the bge-large-en-v1.5 model, which generates an embedding vector denoted as $e^v \in \mathbb{R}^d$. In this case, the output vector dimension is set to $d = 1024$ based on the specifications of the bge-large-en-v1.5 model. This process is formulated as : $\mathbf{E}^{(l)} = f (\mathcal{V})$.  In this way, we can obtain the item embedding matrix $\mathbf{E}^{(l)} \in \mathbb{R}^{|\mathcal{V}|\times d}$.

\subsection{The Adaptive Fast Fourier Transform Powered Mamba Encoder Layer}
\label{sec:adfft}
\textbf{Fast Fourier Transform}. To effectively capture and leverage the inherent periodicities in sequential user behavior like daily, weekly, and other recurring patterns. This work introduces a novel approach that integrates the Fast Fourier Transform (FFT)~\cite{bracewell1989fourier} within a state-space model framework. This addresses a significant limitation of existing sequential recommendation methods, such as IMP~\cite{cheng2024empowering}, which often struggle to accurately discern and disentangle the diverse periodic patterns exhibited by individual users. The computational efficiency of this process is further enhanced through the strategic application of the FFT, denoted here as $\mathcal{F}$. Leveraging the convolution theorem, the FFT facilitates the smoothing of the input sequence $\mathcal{X}_u[t]$ using a carefully designed kernel, $\mathcal{K}$, which acts as a filter to isolate relevant signals while mitigating the effects of noise. The definition of the kernel $\mathcal{K}$ is defined as following:
\begin{equation}
\begin{aligned}
\mathcal{M}(x; \phi)(t) = \int_D \mathcal{K}(t-s; \phi) x_s d_s 
\end{aligned}
\end{equation}
Here $t, s$ denotes time. $\mathcal{M}(x; \phi)$ denotes the kernel integral operator. According to the convolution theorem, the Fourier transform $\mathcal{F}$ is built upon the kernel integral operator, which is expressed as the product of the Fourier transform of the kernel and the Fourier transform of the input signal. Therefore, the process is also shown as follows:
\begin{equation}
\begin{aligned}
\mathcal{M}(x; \phi)(t) = \mathcal{F}^{-1} (\mathcal{W} \cdot \mathcal{F}(x))
\end{aligned}
\end{equation}
Here $\mathcal{F}^{-1}$ denotes the inverse Fourier transform. $\mathcal{W}$ denotes the Fourier transform of the kernel $\mathcal{K}$, and we directly treat $\mathcal{W}$ as a learnable parameter matrix. Then the process of transforming $\mathcal{X}_u[t]$ into frequency domain is mathematically represented as:
\begin{equation}
\begin{aligned}
X_u[k] = \mathcal{F}_u[k] = \sum^{T}_{t=1} \mathcal{X}_u[t] \cdot e^{-j\frac{2\pi}{N}kt}; ~~j=\sqrt{-1}
\end{aligned}
\end{equation}
where $X_u[k] \in \mathbb{C}^{d_f}$ represents the transformed sequence in the frequency domain, with $d_f$ denoting the sequence length in the frequency domain; $k$ is the frequency index; and $j$ represents the imaginary unit.

The integration of the FFT with the Mamba architecture represents a pivotal advancement in sequential recommendation. This synergistic combination yields a powerful enhancement, surpassing the limitations of purely time-domain approaches. The incorporation of the FFT empowers Mamba to not only effectively identify and model the inherent periodic patterns within user sequential data such as daily or weekly cycles but also to mitigate the significant information loss often associated with relying solely on time-domain representations.

\textbf{Adaptive FFT}.
High-frequency components in a user interaction sequence $\mathcal{X}_u$ often represent noise or unwanted artifacts. By filtering out these high frequencies, we can reduce noise and enhance the quality of the user interaction data. The process is shown as follows:
\begin{equation}
\begin{aligned}
X^m_u[k] & = X_u[k] \odot M; 
\end{aligned}
\end{equation}
where $M$ is a binary mask that only retains the frequency that is lower than the $\theta$. Here $\theta$ denotes the threshold of high frequency. Thus, $M$ is obtained via $M :=|F| \leq \theta$ where frequencies below the threshold $\theta$ are retained, while others are filtered out.

\textbf{Inverse Fast Fourier Transform (IFFT)}.
To effectively utilize the frequency-domain information extracted via the FFT and integrate it back into the time-domain representation crucial for sequential recommendation, an IFFT is applied. This transformation converts the frequency-domain features back into a time-domain representation, enabling the model to leverage the insights gained from the frequency analysis within the temporal context of user behavior. The IFFT operation is defined as follows:
\begin{equation}
\begin{aligned}
\label{eq:ifft}
\tilde{\mathcal{X}}_u[t] &= \mathcal{F}^{-1}_u[t]\\& = \frac{1}{T}\sum^{T}_{k=1} \mathcal{W} \cdot X^m_u[k] \cdot e^{j\frac{2\pi}{N}kt}; ~~j=\sqrt{-1}
\end{aligned}
\end{equation}

Here, $\mathcal{F}^{-1}$ denotes the IFFT operation, which maps the frequency-domain features $X^m_u[k]$ to their corresponding time-domain representation $\tilde{\mathcal{X}}_u[t]$.  The term $j$ represents the imaginary unit. The matrix $\mathcal{W} \in \mathbb{C}^{d_f\times d_f}$ is a learned weight matrix that acts as the Fourier transform of the kernel $\mathcal{K}$, effectively modulating the IFFT's reconstruction process. This allows the model to learn a weighted combination of frequency components, refining the time-domain representation and enhancing the accuracy of sequential recommendations by incorporating both temporal and frequency-based insights into user behavior. The importance of this step lies in its ability to bridge the gap between frequency-domain analysis and the time-dependent nature of sequential recommendation.

\textbf{\fmodel\ Block with Adaptive FFT}.
This section details the crucial discretization and processing steps within the Mamba layer, highlighting its contribution to sequential recommendation.  Given the continuous-time parameters $\mathbf{A}$, $\mathbf{B}$, and $\mathbf{C}$ of the state-space model (SSM) and a timestep $\Delta t$, the discretization process transforms these parameters into their discrete-time equivalents, $\bar{\mathbf{A}}$ and $\bar{\mathbf{B}}$,  as follows:

\begin{equation}
\begin{aligned}
\label{eq:mam_dis}
\bar{\mathbf{A}} = \text{exp}(\Delta \mathbf{A}); \quad \bar{\mathbf{B}} = \Delta \mathbf{A}^{-1} \text{exp}(\Delta \mathbf{A}) \cdot \Delta\mathbf{B}
\end{aligned}
\end{equation}
where $\bar{\mathbf{B}}$ ensures accurate discretization.

These discretized parameters are then used within the SSM to process the input sequence $\tilde{\mathcal{X}}_u[t]$, generating a sequence of hidden states $\mathcal{H}$:
\begin{equation}
\begin{aligned}
\label{eq:mam_ssm}
\mathbf{h}_t &= \text{SSM}(\bar{\mathbf{A}}, \bar{\mathbf{B}}, \mathbf{C})(\tilde{\mathcal{X}}_u[t]); \quad \mathcal{H} = \{\mathbf{h}_1, \dots, \mathbf{h}_T\} \\
\tilde{\mathcal{H}} &= \mathcal{H} \odot \text{SiLU}(\text{Linear}(\tilde{\mathcal{X}}_u)) \\
\mathbf{E}^{(m)} &= \text{Linear}(\tilde{\mathcal{H}})
\end{aligned}
\end{equation}
Here, $\odot$ represents the element-wise product, SiLU denotes the SiLU activation function, and Linear represents a linear transformation. The fusion of the hidden states $\mathcal{H}$ with the input features through the element-wise product and SiLU activation enhances the model's ability to capture complex temporal dependencies.  The final linear transformation produces the output embeddings $\mathbf{E}^{(m)}$. This process is critical for sequential recommendation as it allows the Mamba layer to effectively capture both periodic patterns and inherent temporal dynamics within user behavior sequences while simultaneously performing noise reduction.  For complete algorithmic details, please refer to Algorithm~\ref{alg:fmodel}.

\subsection{Fusion Context Embeddings via Gate Mechanism and Model Optimization}
\label{sec:fusion}
In this section, we delve into the pivotal role of the gate mechanism in enhancing sequential recommendation systems. The fusion process, crucial for merging contextual embeddings $\mathbf{E}^{(l)}$ through Long Short-Term Memories (LLMs) and item embeddings $\mathbf{E}^{(m)}$ via the \fmodel\ encoder layer, sets the stage for the introduction of the gate mechanism. This mechanism, encapsulated by the equation below, plays a vital role in refining the fusion process to optimize predictive outcomes for users:
\begin{equation}
\begin{aligned}
\label{eq:gate}
\hat{\mathcal{Y}} = \text{Linear}(\alpha \cdot \mathbf{E}^{(l)} + \beta \cdot \mathbf{E}^{(m)})
\end{aligned}
\end{equation}
Here, $\hat{\mathcal{Y}}$ represents the predicted set of items for users, with $\alpha$ and $\beta$ acting as adaptable parameters. The primary objective is to determine the optimal balance between these parameters to generate the most accurate and personalized item recommendations.
\begin{algorithm}
    \caption{The \textbf{\model} Algorithm}
    \label{alg:model}
    \KwIn{
        $\mathcal{X}_u$: (B, T, $|\mathcal{V}|$), $\mathcal{C}$;
    }
    \KwOut{$\hat{\mathcal{Y}}$: (B, T, 1);\\
    }
    Embedding the item sequence via the embedding layer (Equation~\ref{eq:item_emb});\\
    Represent each item $v_i$ with name, category, and description;\\
    Feed item set into LLMs, $\mathbf{E}^{(l)} \gets \text{LLMs}(\mathcal{C})$;\\
    \For{$e~in~E~epoches$}{
    Obtain $\mathbf{E}^{(m)}$ via $\textbf{\fmodel} $ block;~\textcolor[RGB]{0,0,255}{// Step into \fmodel\ algorithm (Please refer to \fmodel\ algorithm)};\\
    Obtain $\hat{\mathcal{Y}}$ via the Gate mechanism in Equation~\ref{eq:gate};\\
    Calculate loss $\mathcal{L}$ via Equation~\ref{eq:loss};\\
    }
    \textbf{return} $\hat{\mathcal{Y}}: (B, T, 1)$;
\end{algorithm}

\begin{algorithm}[h]
    \caption{The \textbf{\fmodel} Block Algorithm}
    \label{alg:fmodel}
    \KwIn{$\mathcal{X}_u$: (B, T, $|\mathcal{V}|$);\\
    }
    \KwOut{$\mathbf{E}^{(m)}$: (B, T, $|\mathcal{V}|$);\\
    }
    $X_u$[k] = Adaptive FFT($\mathcal{X}_u[t]$);\\
    $X^m_u[k] = X_u[k] \odot M$;\\
    Obtain $\tilde{\mathcal{X}}_u$[t] via IFFT in Equation~\ref{eq:ifft};\\
    \For{$e = 1, 2,..., \fmodel\ \text{layers}$}{
    \textbf{A}: ($|\mathcal{V}|$, d) $\leftarrow$ Parameter;\\
    \textbf{B}: ($|\mathcal{V}|$, T, d) $\leftarrow s_{B}(\tilde{\mathcal{X}}_u)$;\\
    \textbf{C}: (B, T, d) $\leftarrow s_{C}(\tilde{\mathcal{X}}_u)$;\\
    $\Delta$: (B, T, d) $\leftarrow$ $\tau_{\Delta}$(Parameter +  $s_{\Delta}(\tilde{\mathcal{X}}_u)$);\\
    $\bar{\textbf{A}}, \bar{\textbf{B}}$ : (B, L, V, N) $\leftarrow discretize(\Delta, \textbf{A}, \textbf{B})$;\\
    $\mathcal{H}$ $\leftarrow$ SSM ($\bar{\textbf{A}}, \bar{\textbf{B}}, \mathbf{C}$)($\tilde{\mathcal{X}}_u$);\\
    $\tilde{\mathcal{H}}$ $\leftarrow$ $\mathcal{H} \otimes SiLU(Linear(\tilde{\mathcal{X}}_u))$;\\
    $\mathbf{E}^{(m)}$ $\leftarrow$ $Linear(\tilde{\mathcal{H}})$;\\
    }
    \textbf{return} $\mathbf{E}^{(m)}$; 
\end{algorithm}

Moving beyond the fusion process, optimization of \model\ through the use of the cross-entropy loss function further underscores the importance of accurate predictions in sequential recommendation tasks. In this optimization framework, the ground truth item that a user interacts with at a specific time step is denoted by $y^t_u$, while the predicted probability distribution over items at that time step for the user is represented by $\hat{y}^t_u$. The formulation of the loss function, as depicted below, aims to minimize the disparity between the predicted and actual user-item interactions:
\begin{equation}
\begin{aligned}
\label{eq:loss}
\mathcal{L} &=-\frac{1}{|\mathcal{U}|} \sum_{u=1}^{|\mathcal{U}|} \sum_{t=1}^{T} \sum_{v=1}^{|\mathcal{V}|} \Omega  \\
\Omega &= y_{u}^{v,t} \log(\hat{y}_{u}^{v,t}) + (1 - y_{u}^{v,t}) \log(1 - \hat{y}_{u}^{v,t})
\end{aligned}
\end{equation}
Here, the loss function considers the binary indicators $y_{u}^{v,t}$, reflecting user-item interactions, and the corresponding predicted probabilities $\hat{y}_{u}^{v,t},$ with the ultimate aim of refining the model's predictive accuracy and recommendation quality.

Therefore, the gate mechanism serves as a critical component in refining the fusion process of contextual and item embeddings, thereby enhancing the model's predictive capabilities in sequential recommendation systems. Combined with robust optimization strategies like the cross-entropy loss function, these methodologies contribute to the overall effectiveness and performance of recommendation models, ultimately improving user experience and engagement in personalized recommendation scenarios.

\subsection{Model Complexity}
\label{sec:complexity}
This segment delves into the intricacies of our model, \model, by examining its computational complexity, shedding light on the efficiency advantages inherent in our sequential recommendation approach. The foundational Mamba model operates at a computational complexity of $\mathcal{O}(BT|\mathcal{V}|d)$, where $B$ represents the batch size, $T$ signifies the sequence length, $|\mathcal{V}|$ denotes the number of items, and $d$ exemplifies the dimension factor.

Introducing the Fast Fourier Transform into the computational pipeline introduces notable efficiency gains. The Fast Fourier Transform exhibits a time complexity of $\mathcal{O}(BTd \log T)$, while its inverse counterpart operates at $\mathcal{O}(BTd)$, significantly reducing the computational burden compared to the base Mamba model's complexity of $\mathcal{O}(BT|\mathcal{V}|d)$. Consequently, the overall time complexity of our method, \model, remains at $\mathcal{O}(BT|\mathcal{V}|d)$, showcasing a comparable efficiency profile.

Furthermore, to offer comprehensive insights into the operational fabric of our methodology, Algorithm~\ref{alg:model} furnishes a detailed breakdown of the algorithmic processes underpinning \model, cementing its practical utility and computational feasibility in recommendation systems of varying scales of datasets.

%% file: eval.tex
\vspace{-0.15in}
\section{Evaluation}
\label{sec:eval}
In this section, we aim to answer the following research questions:
\begin{compactitem}
    \item \textbf{RQ1.} How does the performance of our model \model\ compare with other state-of-the-art baseline methods?
    \item \textbf{RQ2.} How does our model handle long-term sequences compared to baselines such as BERT4Rec?
    \item \textbf{RQ3.} What is the effect of each component of our method \model\ on performance?
    \item  \textbf{RQ4.} How does the GPU cost and training time of our model \model\ compared to other baselines like SASRec?
    \item \textbf{RQ5.} How robust is our method in terms of mitigating data noise?
    \item \textbf{RQ6.} How do different parameters affect the performance of our model \model?
\end{compactitem}

\begin{table*}[htb]
    \centering
    \caption{Overall performance evaluation across all methods. The best and second best performance are denoted in \textcolor[RGB]{139,0,0}{bold} with red color and \underline{underline} separately. $\ast$ indicates that the best performance is statistically significantly better than the second-best performance ($p \textless $  0.01).}
    \vspace{-0.13in}
    \label{tab:overall_results}
    \resizebox{\linewidth}{!}{
    \setlength{\tabcolsep}{0.9mm}{
    \begin{tabular}{clccccccccccc>{\columncolor{gray!25}}cr}
    \toprule
    Dataset & Metrics &BPR-MF & Caser~ &NARM & GRU4Rec & SASRec & BERT4Rec &LRURec & SR-GNN & FEARec & Mamba4Rec  & RecMamba &\model& \textit{\#Improve}\\
    \midrule
    \multirow{3}{*}{ML-1M} & HR@10 $\uparrow$  &0.1702&0.2399 &0.2735 &0.2934  &0.2977  &0.2958  &0.3057    &0.2997 & 0.2278 &\underline{0.3121}  &0.3097 &\textcolor[RGB]{139,0,0}{\textbf{0.3224}}$^{\ast}$&3.30 \%\\
    ~ & NDCG@10 $\uparrow$&0.0891 &0.1283 &0.1506 &0.1642  &0.1687  &0.1674  &0.1772  &0.1683 & 0.1117   &\underline{0.1822} &0.1805 &\textcolor[RGB]{139,0,0}{\textbf{0.1867}}$^{\ast}$    &2.47 \%\\
    ~ & MRR@10 $\uparrow$ &0.0645&0.0944 &0.1132 &0.1249  &0.1294  &0.1275  &0.1380  &0.1256 & 0.0763   &\underline{0.1425} &0.1389 &\textcolor[RGB]{139,0,0}{\textbf{0.1455}}$^{\ast}$    &4.75\%\\
    \midrule
    \multirow{3}{*}{New York} & HR@10 $\uparrow$ &0.0285&0.0314&0.0336  &0.0355  &0.0360   &0.0367  &0.0392  &0.0402 & 0.0507  &\underline{0.0544} &0.0542 &\textcolor[RGB]{139,0,0}{\textbf{0.0578}}$^{\ast}$  &6.25\%\\ 
    ~ & NDCG@10 $\uparrow$ &0.0154&0.0176&0.0218  &0.0214  &0.0223  &0.0225  &0.0219  &0.0228 & 0.0260   &\underline{0.0278} &0.0265 &\textcolor[RGB]{139,0,0}{\textbf{0.0294}}$^{\ast}$    &5.76\%\\
    ~ & MRR@10 $\uparrow$ &0.0102&0.0124&0.0132  &0.0138  &0.0140  &0.0142  &0.0136  &0.0139 & 0.0186   &\underline{0.0198} &0.0187 &\textcolor[RGB]{139,0,0}{\textbf{0.0210}}$^{\ast}$    &6.06\%\\
    \midrule
    \multirow{3}{*}{California} & HR@10 $\uparrow$ &0.0277&0.0304&0.0324  &0.0326  &0.0335  &0.0331    &0.0340  &0.0338 & 0.0455  &0.0505  &\underline{0.0514}  &\textcolor[RGB]{139,0,0}{\textbf{0.0543}}$^{\ast}$  &5.64\%\\ 
    ~ & NDCG@10 $\uparrow$ &0.0150 &0.0155&0.0162  &0.0168  &0.0171  &0.0169  &0.0174  &0.0173 & 0.0209   &0.0235  &\underline{0.0236}   &\textcolor[RGB]{139,0,0}{\textbf{0.0253}}$^{\ast}$ &7.20\%\\
    ~ & MRR@10 $\uparrow$ &0.0106&0.0115 &0.0128 &0.0132  &0.0136  &0.0132  &0.0136  &0.0138  & 0.0136   &0.0154  &\underline{0.0159}  &\textcolor[RGB]{139,0,0}{\textbf{0.0178}}$^{\ast}$  &11.94\%\\
    \midrule
    \multirow{3}{*}{Texas} &HR@10 $\uparrow$ &0.0280&0.0325 &0.0334 &0.0350  &0.0352   &0.0354  &0.0356  &0.0355 & 0.0422  &\underline{0.0450}&0.0446  &\textcolor[RGB]{139,0,0}{\textbf{0.0472}}$^{\ast}$  & 4.89\%\\ 
    ~ & NDCG@10 $\uparrow$ &0.0147&0.0154&0.0163  &0.0172  &0.0178  &0.0180  &0.0179  &0.0181 & 0.0198   &\underline{0.0203}  &0.0183  &\textcolor[RGB]{139,0,0}{\textbf{0.0225}}$^{\ast}$  & 4.65\%\\
    ~ & MRR@10 $\uparrow$ &0.0082&0.0091&0.0108  &0.0119  &0.0123  &0.0125  &0.0126    &0.0130 &  0.0132  &\underline{0.0140}   &0.0137 &\textcolor[RGB]{139,0,0}{\textbf{0.0152}}$^{\ast}$   & 7.04\%\\
    \bottomrule
    \end{tabular}}
    }
    \vspace{-0.15in}
\end{table*}

\begin{table}
    \centering
    \caption{\textbf{Statistics of the experimented datasets.}}
    \vspace{-0.1in}
    \label{tab:sta}
    \small
    \setlength{\tabcolsep}{1.8mm}{
    \begin{tabular}{ccccccl}
\toprule
\textbf{Datasets} & \textbf{\#Users} & \textbf{\#Items} & \textbf{\#Interactions} & \textbf{Average} \\ \midrule
ML-1M    & 6,040       & 3,416       & 999, 611             & 165.6           \\
New York &6,195       &4,500       &478,903              &106.4          \\
California &7,272       &6,206       &369,469              &50.8          \\
Texas &24,559       &20,375       &1,344,379              &66.0         \\\bottomrule
\end{tabular}}
\end{table}

\subsection{Experiment Settings}
\textbf{Datasets.} Our study utilizes four benchmark datasets covering diverse recommendation scenarios~\cite{yan2023personalized,li2022uctopic}. ML-1M serves as a standard movie recommendation benchmark (165.6 interactions/user). The Amazon Beauty dataset reflects sparse e-commerce behavior (8.9 interactions/user). Three location-aware datasets (New York, California, Texas) demonstrate scaling challenges, with Texas showing the highest engagement (66 interactions/user) and largest volume (27.2M interactions). This selection enables evaluation across: (1) interaction density (8.9-165.6), (2) platform diversity (movies, e-commerce, location-based), and (3) scale variations (198K-27M). Texas's exceptional engagement highlights complex sequential patterns requiring advanced modeling. Table~\ref{tab:sta} compares all four datasets statistically.

\noindent \textbf{Baselines.} We compare our method \model\ with state-of-the-art baselines, including BPR-MF, Caser, NARM, GRU4Rec, SASRec, BERT4Rec, LRURec, SR-GNN, Mamba4Rec and RecMamba. Detailed illustrations are shown as follows:
\begin{compactitem}
    \item \textbf{BPR-MF~\cite{rendle2012bpr}:} BPR-Opt, a versatile Bayesian-derived optimization standard for personalized ranking, is introduced, and a universal BPR learning algorithm using stochastic gradient descent with bootstrap sampling optimizes models according to BPR-Opt, enhancing matrix factorization and adaptive kNN recommenders.

    \item \textbf{NARM~\cite{li2017neural}:} An attention-based blended encoder in NARM captures user intent from sequential actions to create a cohesive session representation.  Recommendation scores are then calculated via bi-linear matching on this representation, with simultaneous learning of item, session, and their correlations.
    \item \textbf{GRU4Rec~\cite{hidasi2015session}:} GRU4Rec proposes an RNN-based approach for session-based recommendations, holistically modeling entire sessions to improve accuracy. It adapts traditional RNNs, including using a ranking loss function, for optimal performance in this specific context.
    \item \textbf{SASRec~\cite{kang2018self}:} SASRec, a self-attention driven sequential model, balances long-term semantic understanding (like RNNs) with focus on a concise set of relevant actions (like Markov Chains).  At each step, it selects relevant past items to predict the next item.
    \item \textbf{BERT4Rec~\cite{sun2019bert4rec}:} BERT4Rec addresses limitations of unidirectional sequential recommendation models by using a bidirectional transformer architecture. To avoid trivial training due to bidirectional context, it employs a Cloze task-predicting masked items-which increases training samples and improves model effectiveness.
    \item \textbf{LRURec~\cite{yue2024linear}}: LRURec uses Linear Recurrent Units for faster inference and incremental processing of sequential inputs in recommendation.  Its simplified linear recurrence operation allows for parallelized training and reduced model complexity.
    \item \textbf{SR-GNN~\cite{wu2019session}:} SR-GNN uses graph neural networks to model session sequences as graphs, capturing complex item transitions and generating more accurate item embeddings than traditional sequential methods.
    \item \textbf{FEARec~\cite{du2023frequency}}: FEARec redesigns conventional self-attention mechanisms to operate in the frequency domain, employing a novel ramp structure that explicitly models both low-frequency (long-term) and high-frequency (short-term) behavioral patterns. This allows the model to distinguish between persistent preferences and transient interactions.
    \item \textbf{Mamba4Rec~\cite{liu2024mamba4rec}:} Mamba4Rec is the first to apply selective State Space Models (SSMs), specifically the hardware-optimized Mamba block, to sequential recommendation. It uses various sequential modeling strategies to improve both effectiveness and inference efficiency.
    \item \textbf{RecMamba~\cite{yang2024uncovering}:} RecMamba uses the Mamba block to selectively model very long (exceeds or equals 2000) user sequences in the context of lifelong recommendation.
\end{compactitem}

\noindent \textbf{Evaluation Metrics.} In alignment with prior research investigations~\cite{liu2024mamba4rec,yang2024uncovering}, we employ a set of established evaluation metrics, comprising Hit Ratio (HR), Normalized Discounted Cumulative Gain (NDCG), and Mean Reciprocal Rank (MRR). These metrics are tailored for evaluation with a truncation threshold set at 10, denoted as HR@10, NDCG@10, and MRR@10, respectively.

\noindent \textbf{Parameter Settings.} In this part, we present detailed illustrations on parameter settings of our method \model\ when our method \model\ achieves the best performance. The batch size of training is set as 2048. The batch size of evaluation is set as 4096. The learning rate is set as $1e-3$. For settings of Mamba, the hidden size is set as 64. The number of layers is set as 1. And the drop ratio is set as 0.2. The loss function is cross entropy. The dimensions of SSM state expansion factor, SSM state expansion factor and local convolution width are set as 32, 4 and 2 respectively.

\subsection{Effectiveness Comparison (RQ1)}
In this section, our primary focus is to meticulously evaluate the efficacy of our model, \model, in contrast to cutting-edge baselines. Meanwhile, performance of Mamba-based methods is better than that of Transformer-based methods, which verifies the effectiveness of Mamba-based structure. The results are meticulously detailed in Table~\ref{tab:overall_results}. Notably, our method excels beyond all other baselines, a feat we attribute to several pivotal factors: 
\textbf{(1)} Adaptive FFT Integration: By implementing adaptive FFT on item sequences, our model \model\ gains a dual advantage. Firstly, it adeptly captures the periodic patterns inherent in user behaviors, enriching its predictive capabilities. Secondly, the adaptive FFT effectively filters high-frequency noise in item sequences, significantly enhancing prediction accuracy.
\textbf{(2)} Leveraging LLMs for Semantic Understanding: Through the strategic utilization of LLMs to model user reviews on items, our model \model\ adeptly captures the nuanced semantics of user behaviors. This integration allows for a profound understanding of user interactions, facilitated by the rich contextual information encapsulated within LLMs. \textbf{(3)} Gate Mechanism Implementation: The incorporation of a gate mechanism plays a pivotal role in determining the optimal trade-off between the embeddings derived from LLMs and those from the \fmodel\ embedding layer. This strategic adoption ensures a harmonious blend of features, further augmenting the model's predictive prowess.

In contrast to Bert-based approaches such as BERT4Rec, a discernible trend emerges showcasing the superior performance of Mamba-based techniques like Mamb4Rec, RecMamba, and our innovative model, \model. This trend underscores the inherent advantages that linear state space models confer upon sequence modeling challenges. Upon comparing our model with Mamba4Rec and RecMamba, it becomes evident that the combination of adaptive FFT, LLMs, and the gate mechanism yields substantial gains in enhancing prediction accuracy.

The substantial performance gap between our method and Frequency-based approaches like FEARec underscores the effectiveness of integrating Mamba with FFT. This combination significantly enhances outcomes compared to traditional Frequency-based methods.

\subsection{Long Sequence Prediction (RQ2)}
\begin{figure}[htb]

\centering
\begin{tabular}{c c}
\\\hspace{-4.0mm}
  \begin{minipage}{0.22\textwidth}
	\includegraphics[width=\textwidth]{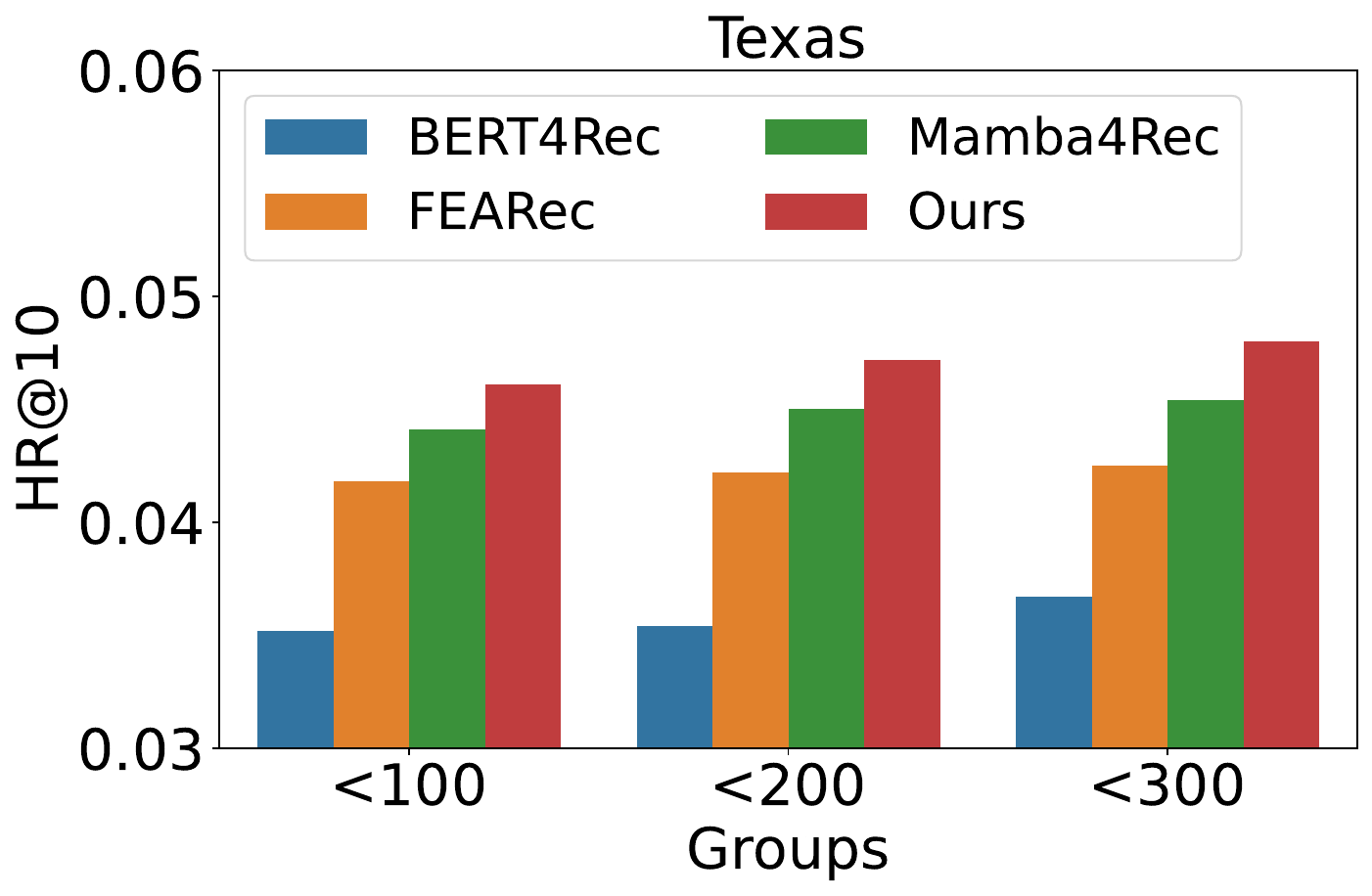}
  \end{minipage}\hspace{-3.mm}
  &
  \begin{minipage}{0.22\textwidth}
	\includegraphics[width=\textwidth]{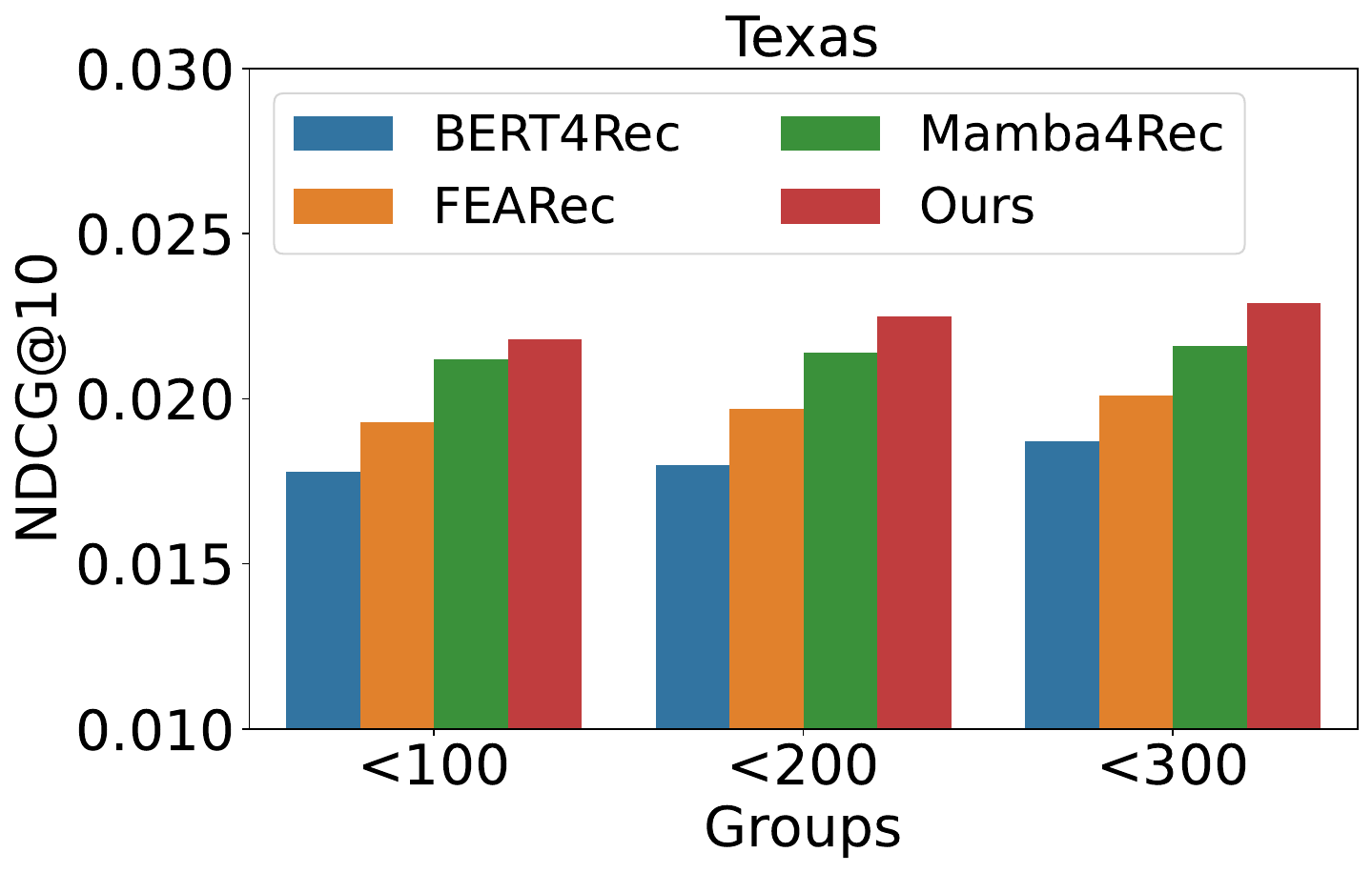}
  \end{minipage}\hspace{-3.0mm}
\end{tabular}
\caption{Performance comparison of baselines on different groups of data}
\label{fig:group_length}
\end{figure}

This section compares the performance of Transformer-based, frequency-based method and Mamba-based recommendation methods on sequences of varying lengths. Specifically, we divide the data into three groups based on sequence length: sequences with less than 100 items, sequences with less than 200 items and sequences with less than 300 items.
As shown in Figure~\ref{fig:group_length}, Mamba-based methods (Mamba4Rec and our proposed method) outperform the Transformer-based method (BERT4Rec), demonstrating the superiority of Mamba architectures for long-sequence prediction. Moreover, our method outperforms Mamba4Rec on long sequences, highlighting its effectiveness in capturing multi-scale periodicities of user behavior in long sequences. Finally, compared with other freqency-based method, our method has more stable performance on long-term prediction period.

\begin{table}[tb]
\centering
    \caption{Performance comparison of different LLMs of \model\ on New York} 
    \label{tab:llm_comp}
\resizebox{\linewidth}{!}{
\begin{tabular}{cccc}
\toprule
Metrics & HR@10 ($\times 1e-2$) & NDCG@10 ($\times 1e-2$) & MRR@10 ($\times 1e-2$) \\ \midrule
BGE-1.5                 &\textcolor[RGB]{139,0,0}{\textbf{5.78}}& \textcolor[RGB]{139,0,0}{\textbf{2.94}}&\textcolor[RGB]{139,0,0}{\textbf{2.10}}     \\
GPT-3.5               &5.68& 2.75&1.97     \\
Llama-3.1 & 5.58 & 2.71 & 1.95 \\ 
\bottomrule
\end{tabular}}
\end{table}

\subsection{Ablation Comparison (RQ3)}
\begin{figure}[htb]
\centering
\begin{tabular}{c c}
\\\hspace{-4.0mm}
  \begin{minipage}{0.22\textwidth}
	\includegraphics[width=\textwidth]{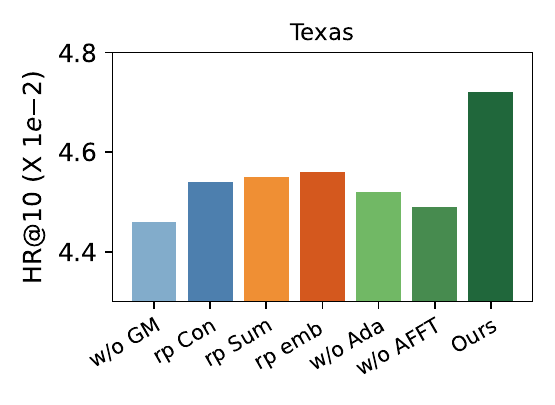}
  \end{minipage}\hspace{-3.mm}
  &
  \begin{minipage}{0.22\textwidth}
	\includegraphics[width=\textwidth]{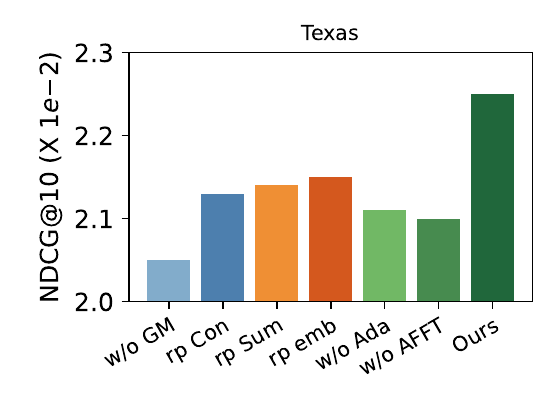}
  \end{minipage}\hspace{-3.0mm}
\end{tabular}
\caption{Ablation study of \model\ on Texas}
\label{fig:ablation_tax}
\end{figure}

This section illustrates the \model\ model's layout and an ablation study on its components. It investigates the impacts of removing or altering key elements such as the gate mechanism and LLMs, providing insights into their effects on performance. The study includes variations like ``w/o GM" removing gate mechanism-generated embedding vectors and ``rp Con" replacing it with a concatenation operation on LLMs and enhanced Mamba layers' vectors. Additionally, the study explores the effects of omitting key algorithmic components like ``w/o Ada" using a standard FFT and ``w/o AFFT" discarding the adaptive FFT, shown in Figure \ref{fig:ablation_tax}.

Our ablation study reveals that each component contributes positively to the overall performance of our model \model. The removal of any single component results in a measurable decrease in effectiveness, underscoring the importance of each element's design. Significantly, the largest performance drop is observed when the gate mechanism (``w/o GM") is removed, strongly suggesting that the integration of LLMs is crucial for effectively capturing the nuanced semantics of user behavior. This finding directly supports our hypothesis regarding the power of LLMs in understanding complex user preferences.  Conversely, the second-worst performing variant is the one lacking adaptive FFT (``w/o AFFT"), emphasizing the critical role of this component in identifying and leveraging the periodic patterns inherent in user behavior. Finally, our experiments replacing the gate mechanism with alternative operations like summation or concatenation demonstrate that the gate mechanism itself is superior to these simpler alternatives, further validating its design and contribution to the model's overall accuracy. 
\vspace*{-0.20in}

To further investigate the influence of different Large Language Models (LLMs) on our model's performance on New York, we conducted experiments using three prominent and widely adopted LLMs: BGE-1.5, GPT-3.5 Turbo and Llama-3.1 7B. The results of these experiments, presented in Figure \ref{tab:llm_comp}, reveal a relatively small performance variation across these different LLMs. This observation suggests that the performance gains are not primarily driven by the specific characteristics of any single LLM. Instead, we attribute this consistent performance across diverse LLMs to the effectiveness of our proposed gate mechanism. The gate mechanism appears to successfully mediate the interaction between the LLM embeddings and the embeddings generated by the enhanced Mamba encoder layer, thereby achieving an optimal balance that minimizes the impact of variations between different LLMs. This consistent performance across different LLMs further underscores the robustness and effectiveness of our gate mechanism in integrating diverse language model outputs.
\vspace{-0.20in}
\begin{figure}[htb]
\centering
\begin{tabular}{c c }
\\\hspace{-4.0mm}
  \begin{minipage}{0.23\textwidth}
	\includegraphics[width=\textwidth]{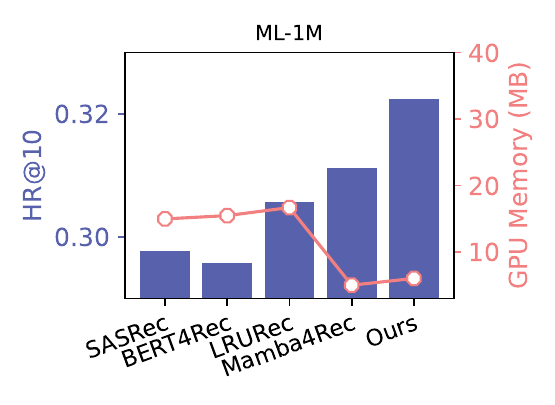}
  \end{minipage}\hspace{-3.mm}
  &
  \begin{minipage}{0.23\textwidth}
	\includegraphics[width=\textwidth]{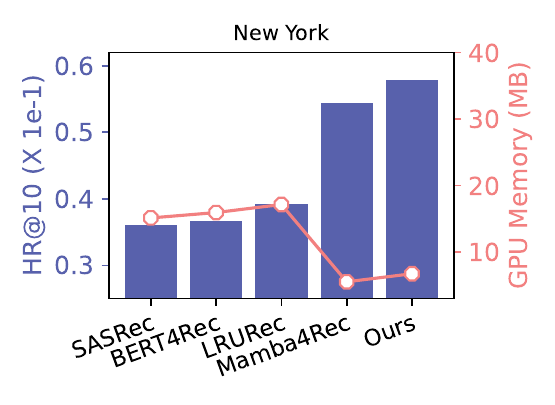}
  \end{minipage}\hspace{-3.0mm}

\end{tabular}
\caption{GPU cost comparison of SASRec, BERT4Rec, LRURec, Mamba4Rec and \model\ on two datasets}
\label{fig:gpu_cost}
\end{figure}

\subsection{Efficiency Comparison (RQ4)}
This section presents a comparative efficiency analysis of our proposed method against several prominent Bert-based approaches. To provide a comprehensive evaluation, we conducted experiments measuring both GPU resource consumption, training time and inference time in terms of each epoch. The results of these experiments are visualized in Figure~\ref{fig:gpu_cost} and Figure~\ref{fig:time_cost_test} respectively, and offer the following key insights:

\vspace{-0.30in}
\begin{figure}[htb]
\centering
\begin{tabular}{c c }
\\\hspace{-4.0mm}
  \begin{minipage}{0.23\textwidth}
	\includegraphics[width=\textwidth]{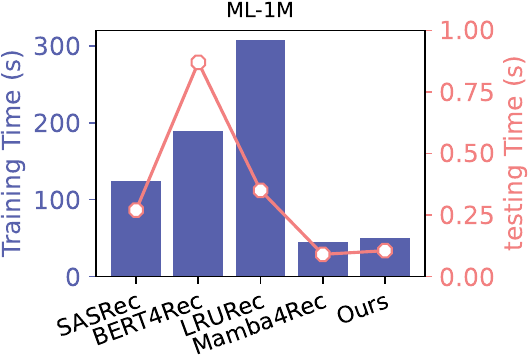}
  \end{minipage}\hspace{-3.mm}
  &
  \begin{minipage}{0.23\textwidth}
	\includegraphics[width=\textwidth]{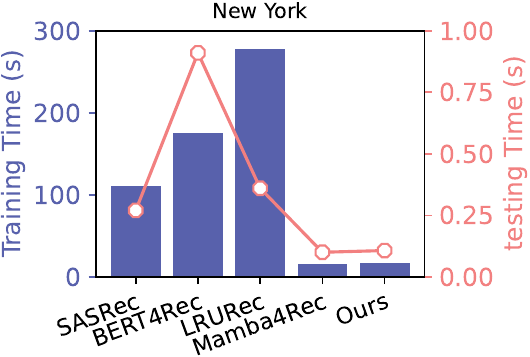}
  \end{minipage}\hspace{-3.0mm}
 
\end{tabular}
\caption{Training time and testing Time comparison of SASRec, BERT4Rec, LRURec, Mamba4Rec and \model}
\label{fig:time_cost_test}
\end{figure}

\textbf{GPU Cost Comparison:} The figure~\ref{fig:gpu_cost} showcases the comparison of GPU memory costs across various methods for the ML-1M and New York datasets. Among the methods such as SASRec, BERT4Rec, LRURec, and Ours, it is evident that Ours outshines the rest in terms of GPU memory efficiency. Specifically, when contrasted with Mamba4Rec, our approach not only matches its efficiency in GPU memory utilization but also outperforms it by delivering superior results based on the HR@10 metric. This highlights the prowess of our method in not only managing GPU memory costs effectively but also achieving elevated performance levels, establishing it as a competitive and impactful solution within the ML-1M and New York dataset contexts.

\textbf{Training and Testing Time Comparison:} Experiments were conducted to evaluate the training time per epoch and the testing time per batch. Figure ~\ref{fig:time_cost_test} shows that our proposed method (\model) demonstrates significantly faster training and testing time compared to several established Bert-based methods, including SASRec, BERT4Rec, and LRURec. This enhanced efficiency stems from two key architectural decisions: the adoption of a Mamba-based architecture, a linear state-space model that inherently requires less computational time than the transformer-based architectures of the Bert-based methods, and the utilization of adaptive FFT with a time complexity of $\mathcal{O}(T log T)$, where $T$ represents the length of the item sequence. While the adaptive FFT contributes to efficiency, the Mamba architecture has a more substantial impact on reducing training time. Notably, our method exhibits similar training and testing times to Mamba4Rec while delivering superior performance, highlighting its practicality and efficiency for large-scale real-life and industry applications.

\vspace{-0.26in}
\begin{figure}[htb]
\centering
\begin{tabular}{c c}
\\\hspace{-4.0mm}
  \begin{minipage}{0.20\textwidth}
	\includegraphics[width=\textwidth, height= 0.8\textwidth]{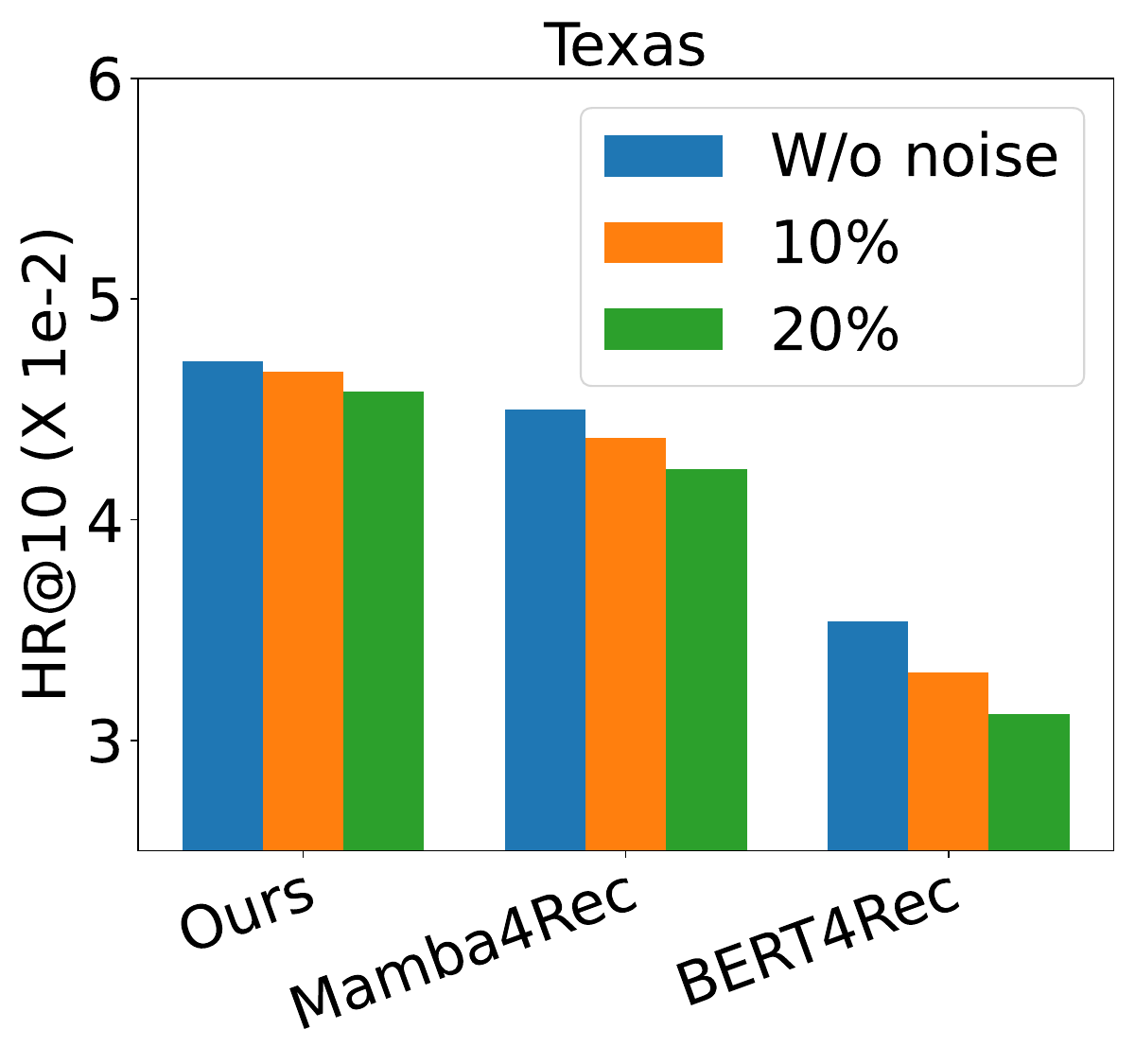}
  \end{minipage}\hspace{-3.mm}
  &
  \begin{minipage}{0.20\textwidth}
	\includegraphics[width=\textwidth, height= 0.8\textwidth]{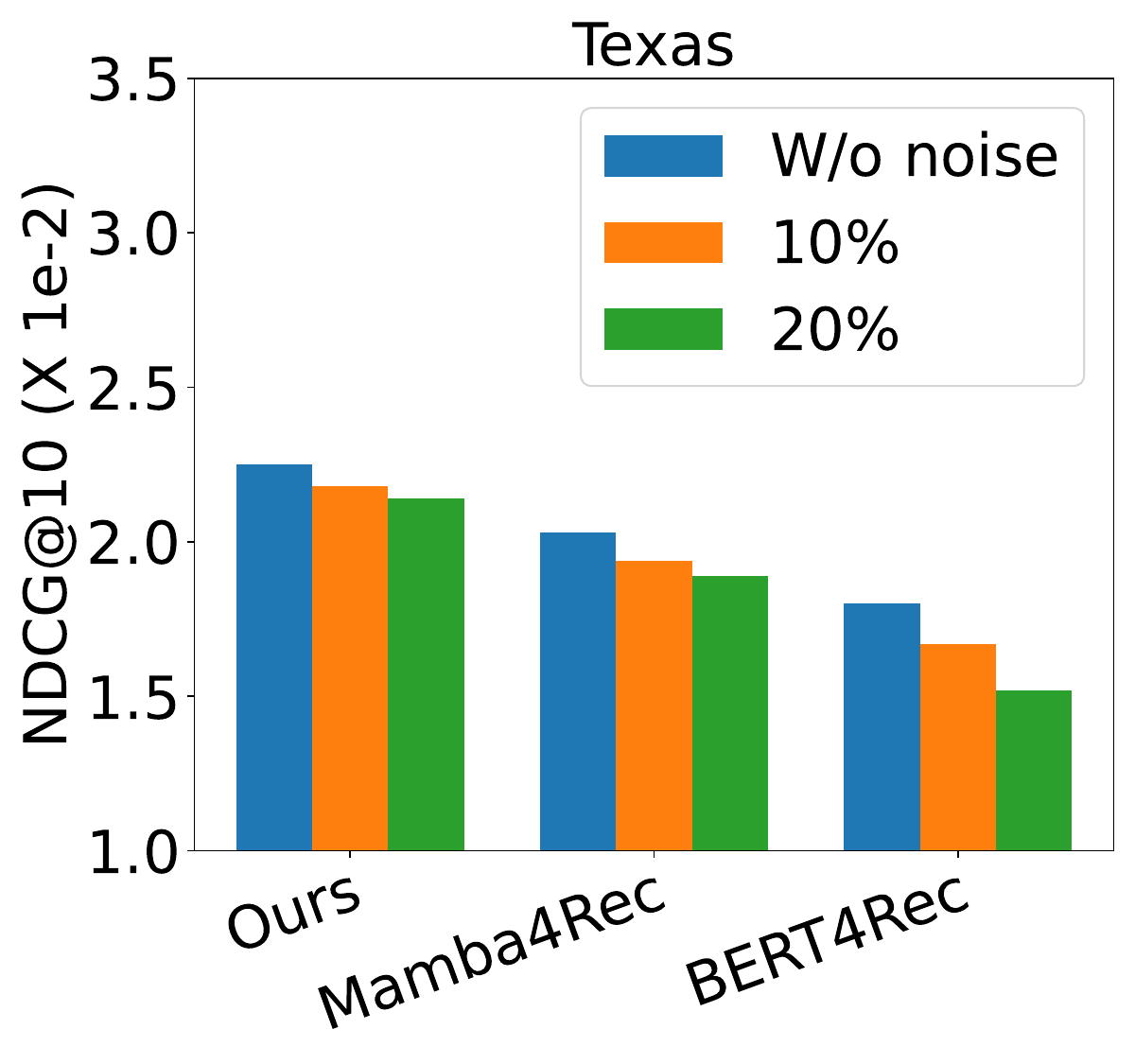}
  \end{minipage}\hspace{-3.0mm}
\end{tabular}
\caption{Performance comparison of robustness}
\label{fig:robustness}
\end{figure}
\subsection{Robustness (RQ5)}
Figure~\ref{fig:robustness} presents a performance comparison of robustness across different models, including Ours, Mamba4Rec, and BERT4Rec, under varying noise levels (0\%, 10\%, and 20\%) on the Texas dataset. Gaussian noise is added to item embeddings during prediction period. The standard deviation of the Gaussian distribution is varied to control the noise level, with settings of 0\%, 10\%, and 20\% of the embedding standard deviation used.

The results show that the Ours model consistently outperforms the other two models, regardless of the noise level. Notably, the performance of ours remains relatively stable, with only a slight decrease as the noise level increases. This suggests that the Ours model is highly robust to noise and can effectively maintain its forecasting accuracy even in the presence of significant noise.

In contrast, both Mamba4Rec and BERT4Rec exhibit a more pronounced decline in performance as the noise level increases. Mamba4Rec, in particular, demonstrates a more gradual decline, suggesting that it is more resilient to noise compared to BERT4Rec. The superior robustness of ours can be attributed to the effectiveness of its adaptive Fourier Transform approach in denoising the input data. This highlights the advantages of the Ours model in real-world applications, where data is often subject to various sources of noise and uncertainty.

\subsection{Hyperparameter Study (RQ6)}
The hyperparameter study focused on the cut ratio in the Fourier Transform and the number of layers in the Mamba model provides valuable insights into performance of recommendation systems.

\textbf{Cut Ratio in Fourier Transform:} The results, presented in Table \ref{tab:hyper}, illustrate that varying the cut ratio in the Fourier transform of \model\ impacts the recommendation metrics. A cut ratio of 0.1 yields the highest HR@10 and NDCG@10 values, indicating that a lower cut ratio captures more high-frequency components effectively. However, the MRR@10 metric is slightly higher at a cut ratio of 0.3, suggesting a trade-off between different evaluation metrics based on the cut ratio chosen. Similarly, at HR@20 and NDCG@20, a cut ratio of 0.1 outperforms others, emphasizing the importance of effectively capturing high-frequency components for top-N recommendations.

\begin{table}[tb]
\centering
    \caption{Hyperparameter study of \model\ on Texas in terms of $\theta$ and mamba layer} 
    \label{tab:hyper}
\begin{tabular}{cccccc}
\toprule
$\theta$ & 0.1 & 0.3 & 0.5 & 0.7 & 0.9 \\ \midrule
HR@10 ($\times 1e-2$)                 &\textcolor[RGB]{139,0,0}{\textbf{4.72}}& 4.69&4.56& 4.58& 4.59     \\
NDCG@10 ($\times 1e-2$)               &\textcolor[RGB]{139,0,0}{\textbf{2.25}}& 2.18& 2.17& 2.18& 2.18     \\
MRR@10 ($\times 1e-2$)               & \textcolor[RGB]{139,0,0}{\textbf{1.52}}& 1.44& 1.45& 1.47& 1.46    \\ \midrule
Mamba Layer  & 1 & 2 & 3 & 4 & 5 \\ \midrule
HR@10 ($\times 1e-2$)                 &\textcolor[RGB]{139,0,0}{\textbf{4.72}}& 4.69& 4.56& 4.58& 4.59    \\
NDCG@10 ($\times 1e-2$)               &\textcolor[RGB]{139,0,0}{\textbf{2.25}}& 2.18& 2.17& 2.18& 2.18     \\
MRR@10 ($\times 1e-2$)               & \textcolor[RGB]{139,0,0}{\textbf{1.52}}& 1.44& 1.45& 1.47& 1.46    \\ \bottomrule
\end{tabular}
\vspace{-0.25in}
\end{table}

\textbf{Number of Layers in Mamba:} The number of layers in the \fmodel\ of \model\ also plays a significant role in the system's performance. The results, presented in Table \ref{tab:hyper}, show that HR@10 and NDCG@10 metrics decrease with an increase in the number of layers, indicating a potential overfitting issue as the model complexity grows. Conversely, MRR@10 remains relatively stable across different layer configurations. For HR@20 and NDCG@20, the performance follows a similar trend, with a decrease in accuracy as the number of layers increases, emphasizing the need to balance model complexity with predictive performance.

%% file: model.tex
\section{Related Work}
\subsection{Sequential Recommendation} In the realm of sequential recommendation research, transformer-based methodologies have garnered significant attention, with early works like BPR-MF~\cite{rendle2012bpr} and Caser~\cite{tang2018personalized} aiming to represent user-item sequences as images, capturing sequential patterns effectively. The evolution to Recurrent Neural Networks (RNNs), exemplified by GRURec~\cite{hidasi2015session}, marked a leap in modeling sequential dependencies within user interactions. LRURec further streamlined this process with rapid inference capabilities and incremental processing. Attention mechanisms, as integrated in NARM~\cite{li2017neural} and subsequent models like SASRec~\cite{kang2018self}, have paved the way for attention-based techniques in sequential recommendation. Additionally, BERT4Rec~\cite{sun2019bert4rec} introduced a bidirectional model via the Cloze task, enhancing predictions with comprehensive context consideration. Despite Transformer-based advancements, they face efficiency challenges due to their attention mechanisms focusing more on individual points. GNNs, such as SR-GNN~\cite{wu2019session}, represent session sequences as graphs, aiding in capturing transition patterns effectively.

Recent advancements~\cite{liu2024mamba4rec,yang2024uncovering,zhang2024matrrec,qu2024ssd4rec,su2024mlsa4rec}, particularly those based on the Mamba architecture~\cite{gu2023mamba}, have emphasized efficiency in handling long sequences. For instance, Mamba4Rec~\cite{liu2024mamba4rec} demonstrates the potential of Mamba for efficient sequential recommendation but falls short in capturing the complex, hierarchical nature of user-item relationships. Similarly, MLSA4Rec~\cite{su2024mlsa4rec} introduces a low-rank decomposed self-attention module with linear complexity, integrating structural bias through Mamba. However, existing methods these approaches remain constrained on periodity caption. Motivated by this, a proposed enhancement integrates Fast Fourier Transform for periodic user pattern capture, noise reduction, and Large Language Models (LLMs) to enrich recommendation accuracy. This proposed fusion approach aims to enhance information fusion among multi-modal sources by leveraging a gate mechanism to integrate insights from LLMs and the enhanced Mamba layer effectively. 

\subsection{Content-enriched Recommendation} The integration of content-enriched recommendations has emerged as a powerful strategy in addressing the cold-start issue within recommendation systems~\cite{lam2008addressing,wu2024survey,hua2023tutorial,zhao2024recommender,jiang2023reformulating,bao2023large,zhao2023recommender}. This approach has gained significant traction alongside the rise of deep learning techniques like RNN or LSTM, enabling the effective incorporation of content features into latent spaces. Wu \textit{et al.}~\cite{wu2022survey} provided existing studies on collaborative recommendations and context-rich recommendations. Notably, in the realm of recommendation systems, there has been a notable focus on the integration of visual-based multimedia elements such as images and videos~\cite{chi2016ubishop,li2021multi}, as well as textual-based content like item descriptions and review texts~\cite{da2021adaptive}. A former study~\cite{jiang2023reformulating} utilizes the semantic comprehension abilities of pre-trained language models. It creates personalized recommendations and connects language models with recommender systems, culminating in recommendations that closely resemble human-generated suggestions. Another former study~\cite{ma2023cross} proposes a recommendation framework called Crossmodal Content Inference and Feature Enrichment Recommendation (CIERec). CIERec utilizes multi-modal data to enhance its performance in cold-start recommendations. A recent study~\cite{cheng2024empowering} proposes to adopt contexts to capture the dynamic, evolving interests of users. Besides, some studies propose to adopt self-attention mechanisms, memory network and so on to capture context information. Different from others, we adopt the Gate Mechanism to combines embeddings of contexts from LLMs and embeddings of correlated item sequential lists to optimize Mamba-based framework and solve the gap between LLMs and Mamba for sequential recommendation applications.

%% file: conclusion.tex
\section{Conclusion}
\label{sec:conclusion}

Sequential recommendation faces challenges due to fluctuating user preferences and complex temporal patterns. While Transformer-based methods struggle with efficiency, Mamba-based approaches offer an alternative but often fail to capture diverse periodic behaviors and mitigate noise.  To address these limitations, we introduce \model, a novel Mamba-based framework incorporating Fast Fourier Transforms (FFTs) and Large Language Models (LLMs). \model\ uses an adaptive FFT mechanism to capture multi-scale periodic user dynamics and suppress noise. A novel gate mechanism fuses LLM embeddings with an enhanced Mamba encoder, leveraging contextual semantics to improve understanding of user preferences.  Comprehensive evaluations across four real-world datasets demonstrate \model's superior performance over existing benchmarks across key metrics, highlighting its effectiveness in addressing the inherent challenges of sequential recommendation.